\documentclass[PRD,reprint,onecolumn,linenumbers,amssymb,amsmath,aps,showpacs,nofootinbib,superscriptaddress]{revtex4}
\usepackage{graphicx}
\usepackage{dcolumn}
\usepackage{bm}

\newcommand{\beq}{\begin{equation}}
\newcommand{\eeq}{\end{equation}}
\newcommand{\beqa}{\begin{eqnarray}}
\newcommand{\eeqa}{\end{eqnarray}}

\begin{document}

\title{AMS-02 positron excess: new bounds on dark matter models and hint for primary electron spectrum hardening}
\affiliation{Key Laboratory of Dark Matter and Space Astronomy, Purple Mountain Observatory, Chinese Academy of Sciences, Nanjing 210008, China}
\affiliation{University of Chinese Academy of Sciences, Beijing, 100012, China}
\author{Lei Feng}
\affiliation{Key Laboratory of Dark Matter and Space Astronomy, Purple Mountain Observatory, Chinese Academy of Sciences, Nanjing 210008, China}
\author{Rui-Zhi Yang}
\affiliation{Key Laboratory of Dark Matter and Space Astronomy, Purple Mountain Observatory, Chinese Academy of Sciences, Nanjing 210008, China}
\affiliation{University of Chinese Academy of Sciences, Beijing, 100012, China}
\author{Hao-Ning He}
\affiliation{Key Laboratory of Dark Matter and Space Astronomy, Purple Mountain Observatory, Chinese Academy of Sciences, Nanjing 210008, China}
\author{Tie-Kuang Dong}
\affiliation{Key Laboratory of Dark Matter and Space Astronomy, Purple Mountain Observatory, Chinese Academy of Sciences, Nanjing 210008, China}
\author{Yi-Zhong Fan$^\ast$}
\affiliation{Key Laboratory of Dark Matter and Space Astronomy, Purple Mountain Observatory, Chinese Academy of Sciences, Nanjing 210008, China}
\author{Jin Chang}
\affiliation{Key Laboratory of Dark Matter and Space Astronomy, Purple Mountain Observatory, Chinese Academy of Sciences, Nanjing 210008, China}
\begin{abstract}
The data collected by ATIC, CREAM and PAMELA
all display remarkable cosmic-ray-nuclei spectrum hardening above the magnetic rigidity $\sim$ 240 GV. One natural speculation is that the primary electron spectrum also gets hardened (possibly at $\sim 80$ GV) and the hardening partly accounts for the electron/positron total spectrum excess discovered by ATIC, HESS and Fermi-LAT. If it is the case, the increasing behavior of the subsequent positron-to-electron ratio will get flattened and the spectrum hardening should be taken into account in the joint fit of the electron/psoitron data otherwise the inferred parameters will be biased. Our joint fits of the latest AMS-02 positron fraction data together with the PAMELA/Fermi-LAT electron/positron spectrum data suggest that the primary electron spectrum hardening is needed in most though not all modelings. The bounds on dark matter models have also been investigated. In the presence of spectrum hardening of primary electrons, the amount of dark-matter-originated electron/positron pairs needed in the modeling is smaller. Even with such a modification, the annihilation channel $\chi\chi \rightarrow \mu^{+}\mu^{-}$ has been tightly constrained by the Fermi-LAT Galactic diffuse emission data. The decay channel $\chi\rightarrow \mu^{+}\mu^{-}$ is found to be viable.
\end{abstract}
\maketitle

\section{Introduction}
In the standard cosmic ray model, most cosmic ray electrons
are from supernova remnants while the cosmic ray positrons are
mainly produced through hadronic processes as cosmic ray protons collide
with intergalactic hydrogen \cite{Ginzburg64book,Nagano00RMP}. In conventional
approach, the injection spectrum of electrons (positrons) is
taken as a single power-law. Since diffusion and electron/positron
cooling are more efficient in higher energies,
the spectrum should soften with energy and the
positron-to-electron ratio (i.e., $\Phi_{e^{+}}/(\Phi_{e^{+}}+\Phi_{e^{-}})$, where $\Phi$ is the flux) should drop with energy
monotonously \cite{Strong07Rev}. Hence there should be no prominent feature at
TeV energies in cosmic ray electron/positron total spectrum, neither in the positron-to-electron ratio.
The situation has changed dramatically since 2008. The Advanced Thin Ionization Calorimeter (ATIC \cite{atic}),
Polar Patrol Balloon-borne Electron Telescope with Scintillating
Fibers (PPB-BETS \cite{ppb-bets}), The High Energy Stereoscopic System (HESS \cite{HESS}) and Fermi Large area Telescope (Fermi-LAT \cite{fermi}) reported
 cosmic ray electron/positron total spectrum up to TeV and found a hardening/bump
in the energy range from 100 GeV to 1 TeV.  Almost simultaneously the Payload for Antimatter Matter Exploration and
Light-nuclei Astrophysics (PAMELA) discovered an unambitious rise of the positron-to-electron
ratio above $\sim 10$ GeV \cite{pamela-positron}. Such a peculiar rising behavior has been confirmed by Fermi-LAT though in an indirect way \cite{Fermi2012}. People call the well-established electron/positron total spectrum hardening/bump and the increasing positron-to-electron ratio, unexpected in the standard model, the excesses or anomalies.
These interesting features draw a lot of attention, and various
physical origins, in particular new astrophysical sources and new physics (dark matter), have been extensively explored \cite[see][for recent reviews]{fan}. Very recently, the AMS-02 collaboration has released their first result on positron fraction in cosmic rays, which confirms the
positron excess with unprecedented accuracy up to the energy $\sim 350$ GeV \cite{AMS-2013}.

On the other hand, the protons and Helium are the most abundant cosmic-ray components, and the spectrum of these cosmic-rays up to the so-called ``knee" can be described by a single power law \cite{Ginzburg64book,Nagano00RMP}. Surprisingly,
the spectra of protons (Helium) measured by ATIC \cite{lab1}, Cosmic Ray Energetics And Mass (CREAM \cite{lab2}) and PAMELA \cite{lab3} show
a remarkable hardening at the magnetic rigidity $\sim$ several hundred GV (GeV/nucleon). We call such a kind of spectrum hardening the cosmic ray nuclei excesses, which challenge the current paradigm of cosmic-ray acceleration and propagation in the Galaxy. Various interpretations have been put forward, such as the spectrum superposition of the local source and the background \cite{Vladimirov2012}, the non-linear acceleration of cosmic rays \cite{Ptuskin2013}, and superposition of the injection spectra of the cosmic ray sources \cite{lab4}. Instead of performing an advanced study of the possible physical origin of the nuclei excesses, in this work we simply take such observational indication (i.e., the hardening of the cosmic ray proton and helium spectra above $\sim 240$ GV \cite{lab3}) as the main motivation to consider the possibility that the primary electron spectrum gets hardened.

This work is structured as follows. In section II we discuss the possibility that the primary electron spectrum gets hardened and then investigate the observational signature. In section III we take the latest AMS-02 positron ratio data to set new bounds on the dark matter models and to constrain the possible primary electron spectrum hardening. The implication on the physical origin of the cosmic ray nuclei excesses is also investigated.
We summarize our results with some discussion in Sec. IV.

\section{Possible spectrum hardening of the primary electrons injected from supernova remnants}\label{sec:II}
The cosmic-ray electron spectrum is connected to the
proton spectrum for two good reasons \cite{Vladimirov2012}: (1) The electrons propagate in the Galaxy
in the same magnetic fields as nuclei; (2) Some, if not all,
CR electrons are produced by the same sources as nuclei. The nucleon injection spectrum has a hardening at a magnetic rigidity $\epsilon_{\rm h,n}\sim 240$ GV, it is thus natural to assume a spectrum hardening of the primary electrons injected from supernova remnants (hereafter we call such primary electrons the {\it background}). In \cite{Vladimirov2012} the authors attributed {\it all} the $e^{-}+e^{+}$ excesses at energies $\gtrsim 100$ GeV   to the background spectrum hardening but found that the positron-to-electron ratio is not possible to reproduce (see their Fig.12 and Fig.13).  We instead suggest that the spectrum hardening of the background electrons just accounts for {\it part of} the $e^{-}+e^{+}$ excesses detected by current instruments. With the given electron/positron total spectrum detected by (for example) Fermi-LAT, {\it the spectrum hardening of the background electrons inevitably leads to the softening of the rest ``additional component" which has been widely assumed to consist of electron and positron pairs. Hence the increasing behavior of positron-to-electron ratio should be shallower than that in the absence of a spectrum hardening.} Such a modification is of our interest since a reliable estimate of the physical parameters of the ``additional component" emitter (from either dark matter annihilation/decay or pulsars) to address the excesses is not achievable if the background spectrum hardening has not been properly taken into account.

The rigidity ($\epsilon_{\rm h,e}$) at which the possible electron spectrum hardening presents is hard to reliably estimate, so is the change of the spectral index of primary electrons ($\delta$). In general, {\it the energy and magnitude of
the hardening do not have to be the same for electrons and protons} since the electron-to-proton ratio may vary with energy and
with source type \citep{Vladimirov2012}. However based on current cosmic ray data one may be able to have some reasonable speculation. For example, the magnitude of the hardening of primary-electron spectrum may be within the range of $E^{0.18}-E^{0.3}$ (i.e., $\delta \sim 0.18-0.3$), as reported for proton (helium) cosmic rays \cite{lab3}.
In the diffusive-reacceleration model, the fit to the cosmic ray data with GALPROP requires a spectrum softening at the magnetic rigidity $\epsilon_{b,n}\sim 11.5$ GV for all nucleons but $\epsilon_{b,e}\sim 4$ GV for electrons \cite{galprop}. Since the injection spectrum of nucleons gets hardened at $\epsilon_{\rm h,n} \sim 240$ GV, the spectrum hardening of the background electrons could present {\it for example} at $\epsilon_{\rm h,e} \sim \epsilon_{\rm h,n} (\epsilon_{\rm b,e}/\epsilon_{\rm b,n})\sim 80$ GV though other values are possible. If the hardening is caused by the injection spectrum of the cosmic rays, the background electron spectrum still gets softened at energies above $\sim 100$ GeV due to the synchrotron and inverse Compton energy losses of high energy electrons. A nearby source, if plays a key role in producing the nuclei excesses, can also give rise to a non-ignorable primary electron spectrum hardening since the cooling during the travel is far more crucial in modifying the spectrum for electrons than the nuclei. 

Throughout this work we assume that the hardening of the background spectrum takes place only once. At least in principle multiple spectrum hardening is possible. The AMS-02 data to be released in a few months can shed valuable light on such a kind of possibility.

\section{Positron ratio data together with the electron/positron spectrum data: Bounds on models}
Dark matter is a form of matter necessary to account for gravitational effects observed in very large scale structures such as the flat rotation curves of galaxies and the gravitational lensing of light by galaxy clusters that cannot be accounted for by the amount of observed/normal matter \cite{lsp}. The most widely discussed candidate is the so-called weakly interacting massive particles (WIMPs), which may annihilate with each other or decay and then produce particle pairs such as photons, electrons and positrons and so on \cite{lsp}. That is why dark matter may be able to account for the observed positron and electron excesses, as extensively examined in the literature \cite{Yin2009,Decay2009,Decay2010,Cirelli2012}.

Alternatively, the positron excess detected by PAMELA/Fermi-LAT/AMS-02 may be mainly contributed by pulsars. High energy electrons/positrons may be generated through the cascade of electrons accelerated in the magnetosphere of pulsars
\cite{zhangli,Profum2009}. The energy spectrum of $e^+e^-$ injected to
the galaxy from pulsars can be parameterized as a broken power-law
with the cutoff at $E_c$, i.e., ${\rm d}N/{\rm d}E \propto A_{\rm psr}
E^{-\alpha}\exp(-E/E_c)$, where $E_c$ ranges from several tens GeV
to higher than TeV and the power-law index $\alpha$
ranges from 1 to 2.2 depending on the gamma-ray and radio
observations \cite{zhangli,Profum2009,Malyshev2009}.
The spatial distribution
of pulsars is parameterized as \cite{distrib}
\begin{equation}
f(R,z)\propto\left(\frac{R}{R_{\odot}}\right)^{2.35}\exp\left[-\frac{5.56(R-
R_{\odot})}{R_{\odot}}\right]\exp\left(-\frac{|z|}{z_s}\right),
\end{equation}
where $R_{\odot}=8.5$ kpc is the distance of solar system from the
Galactic center, and $z_s\approx 0.2$ kpc is the scale height of the
pulsar distribution. The other parameter
appearing in the above equation is the normalization factor $A_{\rm
psr}$ that will be determined in our data fit.

In the latest modeling the most widely adopted data include the PAMELA electron spectrum \cite{Pamela2011}, the Fermi-LAT electron + positron spectrum \cite{fermi}, and the AMS-02 positron fraction data \cite{AMS-2013}. Since these data were collected by different instruments, the systematic errors are hard to estimate. At energies $\leq 30$ GeV, the Fermi-LAT {\it standard} electron + positron flux (i.e., the ``standard high energy selection data") reported in \cite{fermi} is below the PAMELA electron flux \cite{Pamela2011}. In a recent paper to estimate the positron-to-electron ratio, with the updated instrument response functions taking into account ``ghost events" the Fermi-LAT collaboration also presented the electron/positron total spectrum in the energy range $20-200$ GeV \cite{Fermi2012}, which seems to be more consistent with the PAMELA electron spectrum. At energies $\geq 200$ GeV, the difference of the electron + positron spectra between Fermi-LAT and ATIC might be partly due to the very different path length of the detectors.  To check such a possibility the Fermi-LAT collaboration has performed a dedicated analysis in which they selected events with the longest path
lengths in the calorimeter. Two additional requirements for these events are that
 they do not cross any of the boundary gaps between calorimeter tower modules and they have sufficient track length in the tracker for a good
direction reconstruction \cite{fermi}. For such an event sample, the spectrum seems to be a bit harder than that obtained in the standard high energy selection (see Fig.19 of \cite{fermi} for a comparison) though the consistence is well in view of the relatively large systematic errors. Considering these facts, in this work we model two sets of data separately, including (I) The Fermi-LAT {\it standard} electron + positron flux data \cite{fermi} and the AMS-02 positron fraction data  \cite{AMS-2013}; (II) the Fermi-LAT long path electron + positron spectrum data presented in \cite{fermi}, the updated Fermi-LAT electron + positron spectrum in the energy range $20-200$ GeV \cite{Fermi2012}, the PAMELA electron spectrum data \cite{Pamela2011}, and the AMS-02 positron fraction data.

As already mentioned before, the electron excesses could originate from either dark matter annihilation/decay or pulsars and the background electrons  might display significant spectrum hardening. We try to jointly fit the electron/positron total spectrum and the positron-ratio data in the following scenarios, including (a) background without spectrum hardening + dark matter annihilation/decay into $\mu^{+}\mu^{-}$; (b) background without spectrum hardening + pulsars; (c) background with spectrum hardening + dark matter annihilation/decay into $\mu^{+}\mu^{-}$; (d) background with spectrum hardening + pulsars. For the second set of data, we will also fit the data in the model of ''background with spectrum hardening + pulsars + dark matter annihilation into $e^{+}e^{-}$" (i.e., scenario (e)).

The electrons traveling in the Galaxy suffer from inverse Compton scattering of interstellar background photons (e.g., cosmic microwave background, dust emission and star light) and boost these photons to GeV energies, becoming part of the Galactic diffuse emission. Hence the Galactic diffuse emission detected by space telescopes can be used to reliably constrain the physical parameters of dark matter particles \cite{lsp}. The latest bounds set by the Fermi-LAT Galactic diffuse emission data are presented in \cite{Fermi2012ApJ}. To make use of such bounds, our dark matter density distribution profiles as well as the cosmic ray diffusion parameters are taken to be the same as that adopted in \cite{Fermi2012ApJ}.
The smooth dark matter density $\rho$ are parameterized with a Navarro-Frenk-White (NFW) spatial profile \cite{NFW}
\begin{equation}
\rho(r)=\rho_0 (1+R_\odot/r_{\rm s})^{2}/[(1+r/r_{\rm s})^{2}r/R_\odot],
\end{equation}
and the isothermal-sphere (ISO) profile \cite{Isothermal}
\begin{equation}
\rho(r)=\rho_0 (R_\odot^{2}+r_{\rm c}^{2})/(r^{2}+r_{\rm c}^{2}),
\end{equation}
respectively. Where $\rho_0=0.43~{\rm GeV~cm^{-3}}$ is the local density of dark matter,
$r_{\rm s}=20$ kpc and $r_{\rm c}=2.8$ kpc are two scale radii. Throughout this work, NFW (ISO) profile is adopted to probe the dark matter decay (annihilation) scenario. As a result of the significant cooling of the high energy electrons, only the relatively nearby electrons/positrons can reach the Earth. Hence the result of modeling electron/positron data does not depend on the dark matter distribution profile sensitively. However, very tight constraint on the dark matter annihilation scenario has been set by the Galactic diffuse emission and the NFW profile is strongly disfavored.

We adopt the GALPROP \cite{galprop} package to numerically calculate the propagation of the cosmic ray particles, including both those from astrophysical sources and the contribution from dark matter annihilation or decay. The cosmic ray diffusion parameters taken into account are the Halo height $z_{\rm h}=4$ kpc, the diffusion coefficient $D_0=5.3 \times 10^{28}~{\rm cm^{2}~s^{-1}}$, the diffusion index $\delta=1/3$, the Alfven velocity $V_{\rm A}=33.5~{\rm km~s^{-1}}$, the nucleon injection indexes below and above the break rigidity $\rho_{\rm b,n}=11.5$ GV are $1.88$ and $2.39$, respectively. To reasonably fit the data, two codes have been developed. One is based on the MINUIT (http://seal.web.cern.ch/seal/work-packages/mathlibs/minuit/) and the other is based on the COSMOMC (http://cosmologist.info/cosmomc/). In most cases the former can provide a very-quick and reasonable (i.e., $\chi^2/{\rm d.o.f}\leq 1$, where d.o.f is the degree of the freedom) fit of the data. However the obtained fit parameters are usually not the ``best" and reliable estimates of the parameter spaces are not achievable. The latter can provide us the best fit parameters but is very time-consuming. The main purpose of this work is to investigate whether the proposed scenarios can reasonably reproduce the data or not, therefore in some modelings we take the MINUIT-based code. The COSMOMC-based code is adopted if we want to be sure that a reasonable fit is un-achievable or aim to reliably estimate the parameter space.

\begin{table}[!htb]
\begin{small}
\caption {The fit parameters of the first set (I) of electron/positron cosmic ray data.}
\begin{tabular}{ccccccccccc}
\hline \hline
scenario & $m_\chi$ & $<\sigma v>$ & $\tau_{\chi \rightarrow \mu^{+}\mu^{-}}$ & $e^-$ injection &$\epsilon_{\rm h,e}$ & $\alpha$ & $E_{\rm c}$ & $c_{e^{+}}$ & $\chi^2/{\rm d.o.f}$ \\
   &  (GeV) & ($10^{-26}$ cm$^3$ s$^{-1}$) & ($10^{26}~{\rm s}$)  & $\gamma_1/\gamma_2$  & (GeV)& & (GeV) \\
\hline
(a) & 1073.5 & 690.7 &  &  2.645/2.645 &  &  & & 1.41 & 252.4/86 & \\
    & 1802.9 &  & 3.62 &    2.637/2.637 & & &  & 1.476 & 333.5/86 & \\
(b) &  &  &  & 2.520/2.520  &  & 1.337  & 651 & 0.812 & 77.6/86 &  \\
(c) & 627.1  & 225.0  &  &  2.797/2.375 & 68.2  & &  & 1.554 & 74.2/84 &  \\
    & 1011.6   &  & 6.96 &  2.792/2.301 & 81.9  &  &  & 1.595  & 92.8/84 &  \\
(d) &  &  &  & 2.678/2.475  & 62.8 & 1.303  & 755.7 & 1.242 & 50.1/83 & \\
  \hline
  \hline
\end{tabular}
\label{table1}\\
\end{small}
\end{table}

\begin{table}[!htb]
\begin{small}
\caption {The fit parameters of the second set (II) of electron/positron cosmic ray data.}
\begin{tabular}{ccccccccccc}
\hline \hline
scenario & $m_\chi$ & $<\sigma v>$ & $\tau_{\chi \rightarrow \mu^{+}\mu^{-}}$ & $e^-$ injection &$\epsilon_{\rm h,e}$ & $\alpha$ & $E_{\rm c}$ & $c_{e^{+}}$ & $\chi^2/{\rm d.o.f}$ \\
   &  (GeV) & ($10^{-26}$ cm$^3$ s$^{-1}$) & ($10^{26}~{\rm s}$)  & $\gamma_1/\gamma_2$  & (GeV)& & (GeV) \\
\hline
(a) & 1694.3  & 1452.1 &  & 2.70/2.70  &  &  &  & 1.619 & 271/87 \\
    & 3048.9 &  & 2.747 &    2.716/2.716 & & & & 1.738 & 337.6/87 \\
(b) &  &  &  & 2.588/2.588  &  & 1.331  & 772.3 & 1.002 & 143.5/86 & \\
(c) & 1176.3 & 575 &  & 2.862/2.441  & 62.1  & &  & 1.717 & 75.0/85 & \\
    & 2139.7 &  &  4.648 & 2.885/2.445  & 55.7  & &  & 1.756 & 89.7/85 & \\
(d) &  &  &  & 2.687/2.333  & 103.9 & 1.338  & 500.0 & 1.091 & 57.6/84 & \\
(e) & 104.7 & 0.55 &  &  2.786/2.445 & 67.3  & 1.2 & 767.5 & 1.547 & 52.2/82 & \\
  \hline
  \hline
\end{tabular}
\label{table2}\\
\end{small}
\end{table}

\begin{figure}
\includegraphics[width=85mm,angle=0]{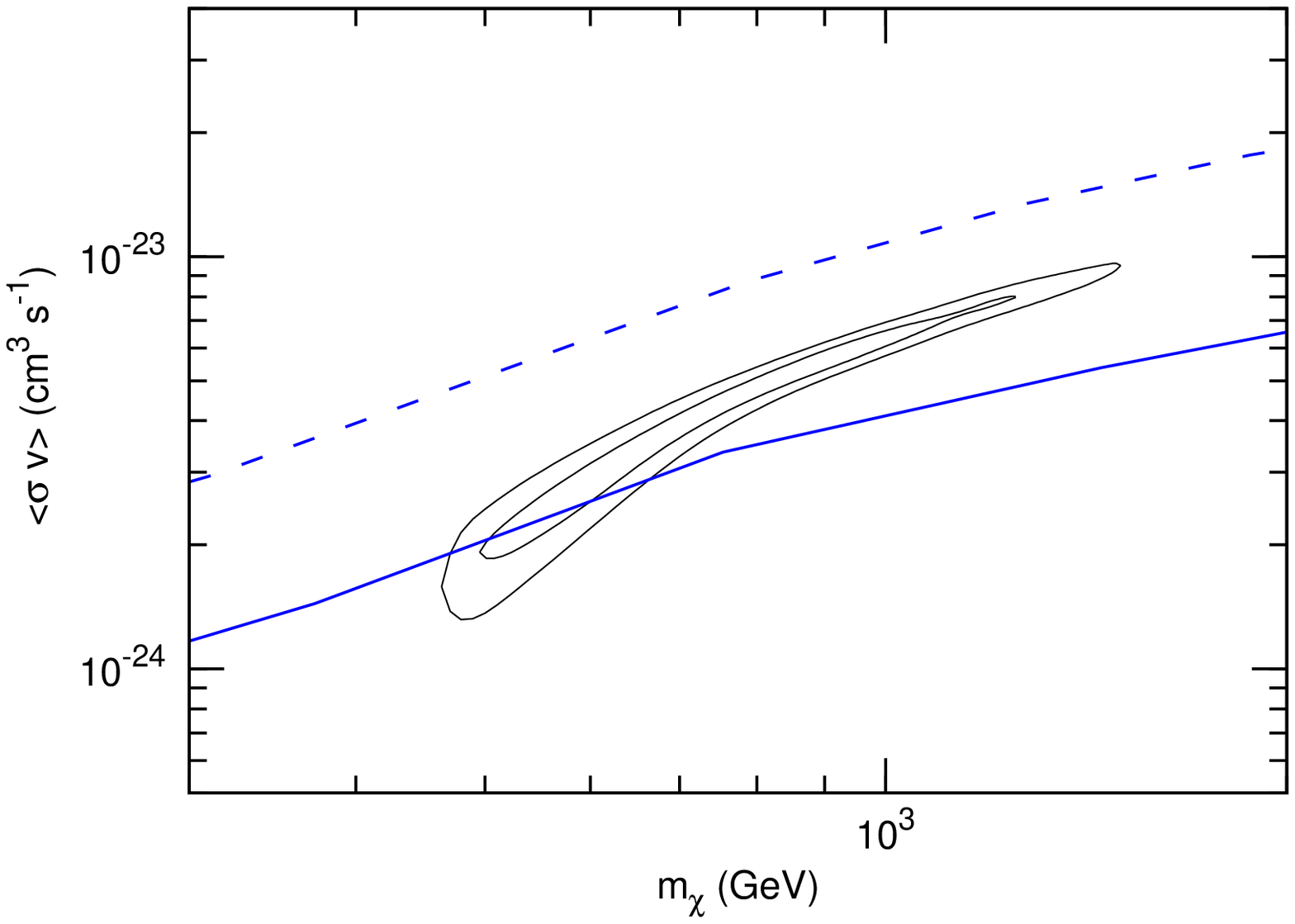}
\includegraphics[width=85mm,angle=0]{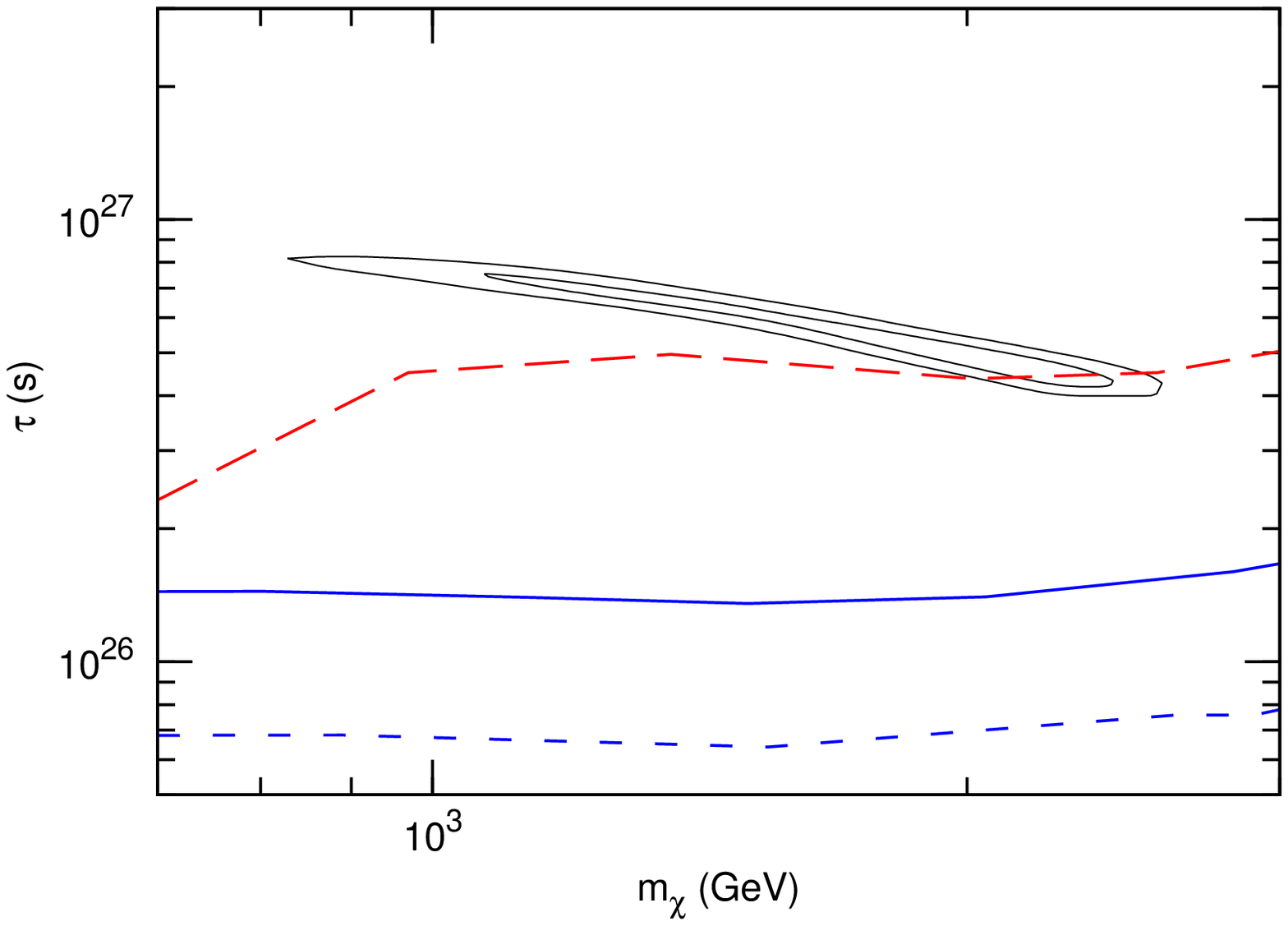}
  \caption{The regions of parameter space ($68.3\%$ and $99.5\%$ confidence levels) which provide a reasonable fit to the second set of data  comparing with the bounds set by Fermi-LAT Galactic diffuse emission (adopted from \cite{Fermi2012ApJ}, the short-dashed line represents the upper limits on $\langle\sigma v\rangle$ found in the analysis with no model of the
astrophysical background, while the solid line is the bound found in the analysis with a modeling of the background) and by the extra-galactic diffuse emission (adopted from \cite{Cirelli2012}, the long-dashed line in the right panel).
Left panel is for the annihilation channel $\chi\chi\rightarrow \mu^{+}\mu^{-}$ and the isothermal-sphere like dark matter distribution model is adopted.
Right panel is for the decay channel $\chi \rightarrow \mu^{+}\mu^{-}$ and the NFW dark matter distribution model is adopted.
In this work we do not present the cases of annhilation/decay into $\tau^{+}\tau^{-}$ since they have been excluded by the $\gamma$-ray observations.}
  \label{fig:1}
\end{figure}

The fit parameters of these two sets (I and II) of electron/positron cosmic ray data are summarized in Tab.\ref{table1} and Tab.\ref{table2}, respectively. To minimize the
effect of solar modulation, we use the data with energies greater
than 10 GeV for $\chi^{2}$ calculation. One can see that both sets of data can not be reasonably fitted within scenario (a), i.e., background without spectrum hardening + dark matter annihilation/decay into $\mu^{+}\mu^{-}$ (see also \cite{Yuan2013,Jin2013nta,Hooper2013}). Interestingly, the first set of data can be reasonably fitted within scenario (b), i.e.,
background without spectrum hardening + pulsars (see also \cite{pulsar2013}), while the second set of data can not be. Such a difference reflects the divergency between these two groups of electron + positron spectrum data. The fits of the data can be considerably improved in the presence of  spectrum hardening of the background, in agreement with \cite{Hooper2013,Yuan2013b}. For simplicity, the spectrum hardening of the primary electrons has been assumed to be the same for all the sources. The inferred $\epsilon_{\rm h,e}$ ($\delta=\gamma_2-\gamma_1$) is in the range of $55-104$ GV ($0.2-0.5$), roughly consistent with the speculations made in section \ref{sec:II}. Such consistences are encouraging, however we would like to caution that the data used in modeling are from different instruments and may suffer from significant systematic errors. The AMS-02 electron/positron spectrum will test the hardening hypothesis soon. We are aware that some models such as multiple pulsars \cite{Yin2013} and the decaying asymmetric dark matter \cite{FengKang2013} can also nicely fit the data. The spectrum hardening of both primary-electrons and nuclei, if confirmed by AMS-02 in the future, will suggest some ``nearby" supernova-remnant-like sources within a radius $R \sim 2.4~{\rm
kpc}~({D_0 \over 10^{28.7}~{\rm cm^{2}~s^{-1}}})^{1/2}({u_{\rm
tot}\over  1~{\rm eV~cm^{-3}}})^{-1/2}({\epsilon_{\rm h} \over
100~{\rm GeV}})^{-1/3}$, where $u_{\rm tot}=u_{\rm B}+u_{\rm cmb}+u_{\rm dust}+u_{\rm
star}$, $u_{\rm B}=B_{\rm IG}^{2}/8\pi$ is the magnetic field energy
density, $u_{\rm cmb}$, $u_{\rm dust}$ and
$u_{\rm star}$ are the photon energy densities of cosmic
microwave background, dust emission and the star emission, respectively. In other words some sources are relatively nearby. The candidate supernova remnants include for example Geminga and Loop I \cite{Shaviv2009,Profum2009}. Here we do not consider nearby pulsars since they are expected to produce electron/positron pairs rather than mainly electrons. The nearby supernova remnants instead produce ``ignorable" amount of positrons. This can be straightforwardly understood as the following. Let us make the``optimistic" estimate, i.e., assuming that the total cosmic ray protons above $\sim 240$ GeV are dominated by the nearby source. The chance for one high energy proton to produce one positron is $P \sim \sigma_{\rm pp}n c \tau/3 \sim 3.5\times 10^{-3}~(n/0.5~{\rm cm^{-3}})(\tau/10^{13}~{\rm s})$, where $\sigma_{\rm pp} \approx 70$ mb is the total cross section of production of pions in proton-proton collision, and $n \sim 0.1-1~{\rm cm}^{-3}$ is the number density of the local interstellar medium. The high energy proton loses about 20\% energy
in one proton-proton collision, roughly one quarter
 converts into positron via the decay of the positively charged pion (i.e.,
$\pi^{+} \rightarrow \mu^{+} + \nu_\mu\rightarrow e^{+} + \nu_{\rm e}+\bar\nu_{\mu}+\nu_{\mu}$). Hence at the energy of $240$ GeV, the positron-to-proton ratio is $\sim P/20^{1.7} \sim 2\times 10^{-5}~(n/0.5~{\rm cm^{-3}})(\tau/10^{13}~{\rm s})$, well below the value $\sim 3\times 10^{-4}$ inferred from the PAMELA/AMS-02 data \cite{pamela-positron,AMS-2013}, where the $E^{-2.7}$-like proton spectrum has been taken into account.

\begin{figure}
\includegraphics[width=60mm,angle=270]{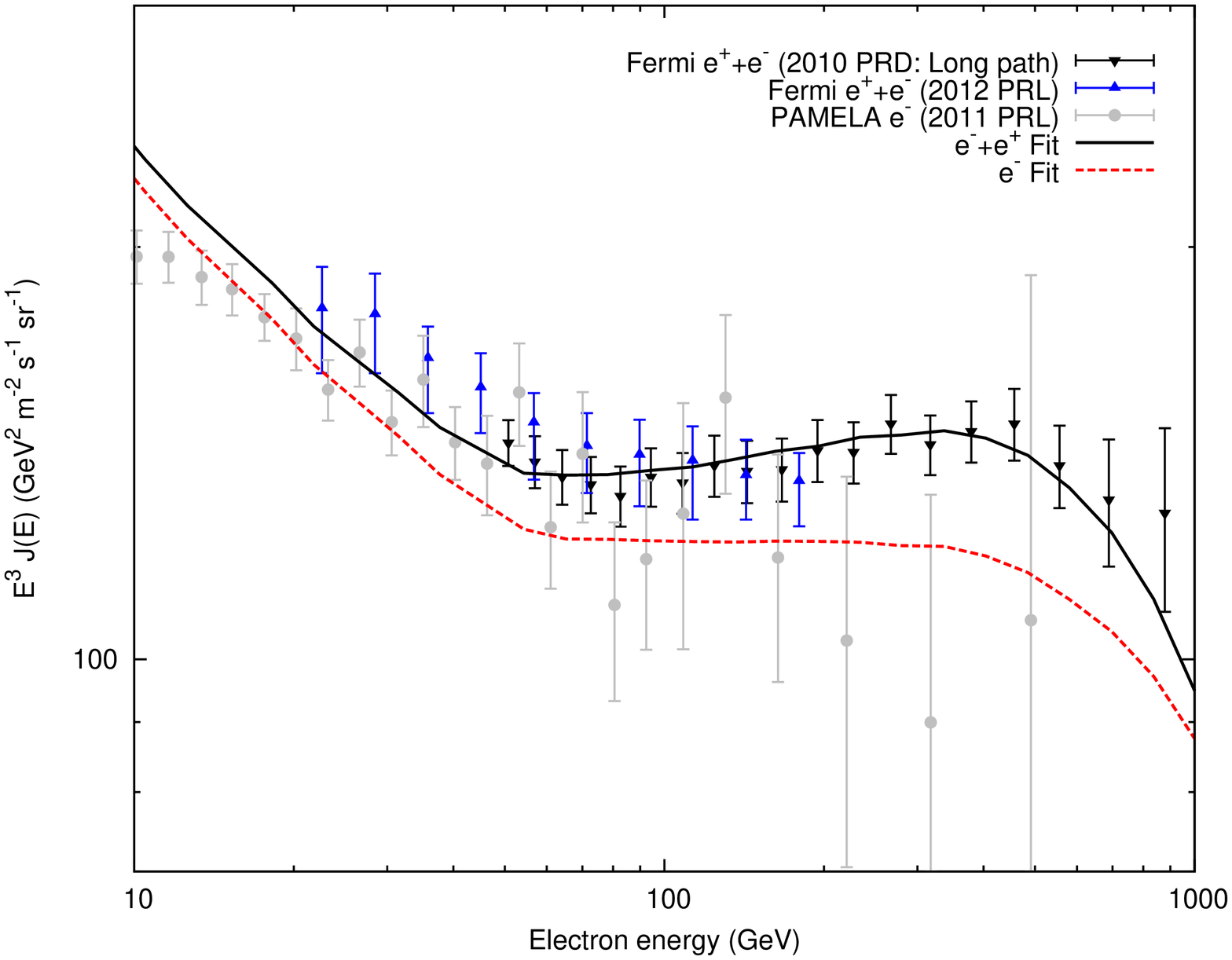}
\includegraphics[width=60mm,angle=270]{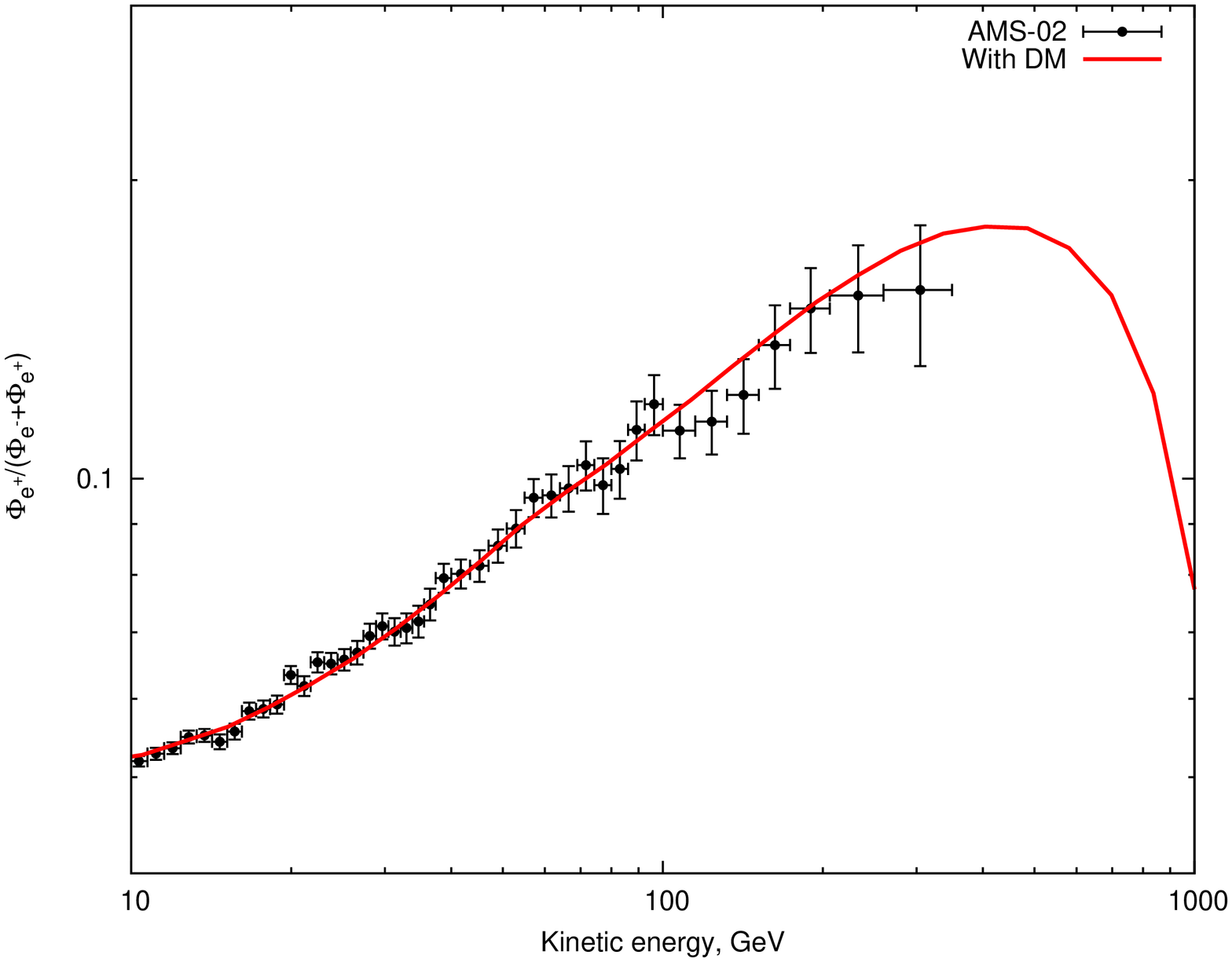}
\includegraphics[width=60mm,angle=270]{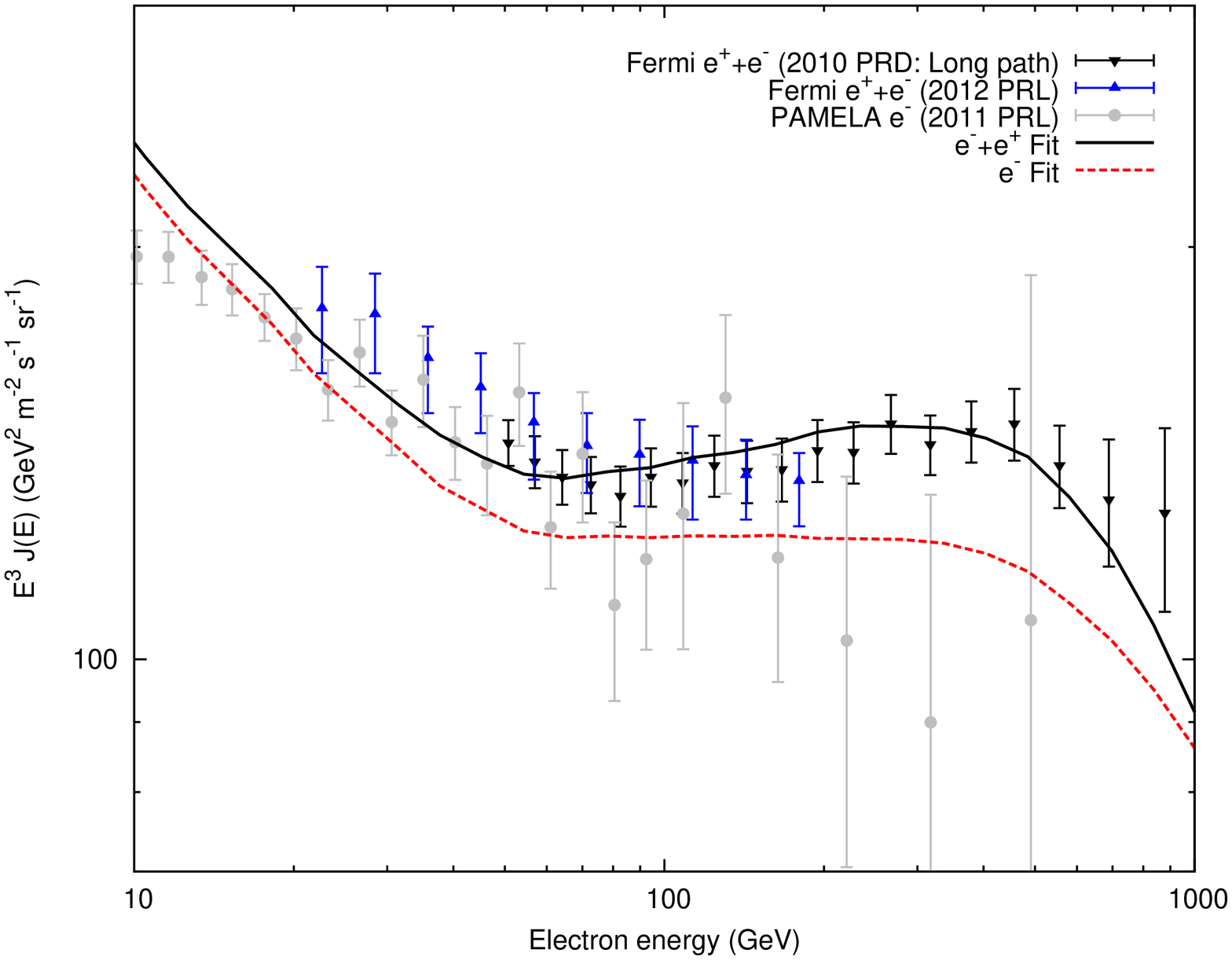}
\includegraphics[width=60mm,angle=270]{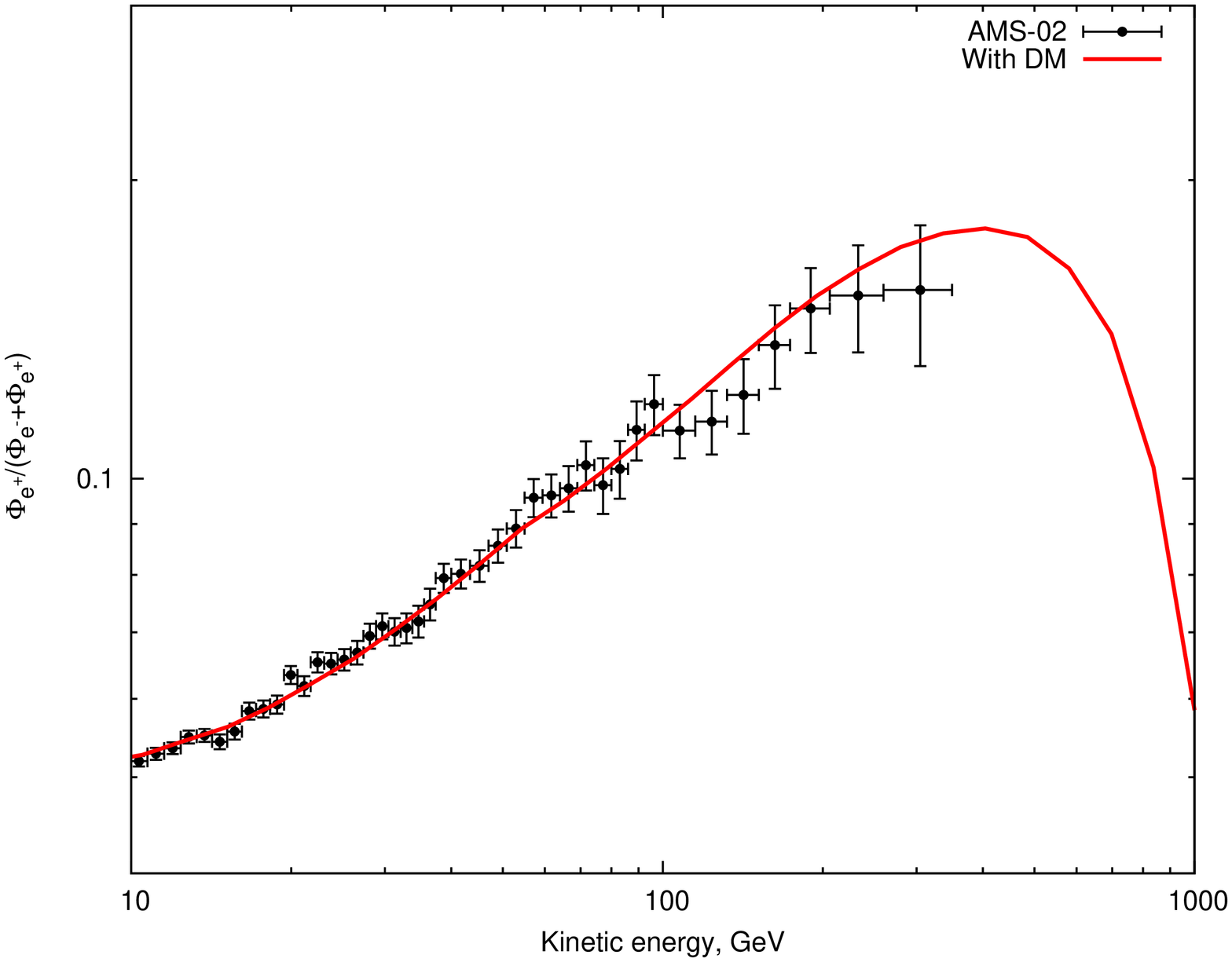}
  \caption{The upper two panels: the ``background with spectrum hardening + dark matter annihilation into $\mu^{+}\mu^{-}$" model for the second set of electron/positron data. The lower panels: the ``background with spectrum hardening + dark matter decay into $\mu^{+}\mu^{-}$" model for the second set of electron/positron data. The best fit parameters are summarized in Tab.II.}
  \label{fig:2}
\end{figure}

In Fig.\ref{fig:1} we present the parameter space allowed by the second set of data in scenario (c). In comparison with the previous fit results for Fermi-LAT electron/positron spectrum and the PAMELA positron ratio data (see e.g. Fig.2 of \cite{Cirelli2012}), our best fit of $<\sigma v>$ ($\tau_{\chi\rightarrow \mu^{+}\mu^{-}}$) is a few times smaller (larger). Such a difference is due to the presence of spectrum hardening of primary electrons, with which the dark-matter-originated electron/positron pair component needed in the modeling is smaller.
Though the fits of the electron/positron data are well (see Fig.\ref{fig:2}), the Galactic diffuse  $\gamma$-ray emission \cite{Fermi2012ApJ} as well as the extra-galactic diffuse $\gamma$-ray emission \cite{Cirelli2012} impose constraints on the models. Evidently, most of the allowed parameter space for $\chi\chi \rightarrow \mu^{+}\mu^{-}$ has been excluded by the Galactic diffuse $\gamma$-ray emission even in the ISO dark matter distribution model while the channel $\chi\rightarrow \mu^{+}\mu^{-}$ is still viable.

\begin{figure}
\includegraphics[width=65mm,angle=270]{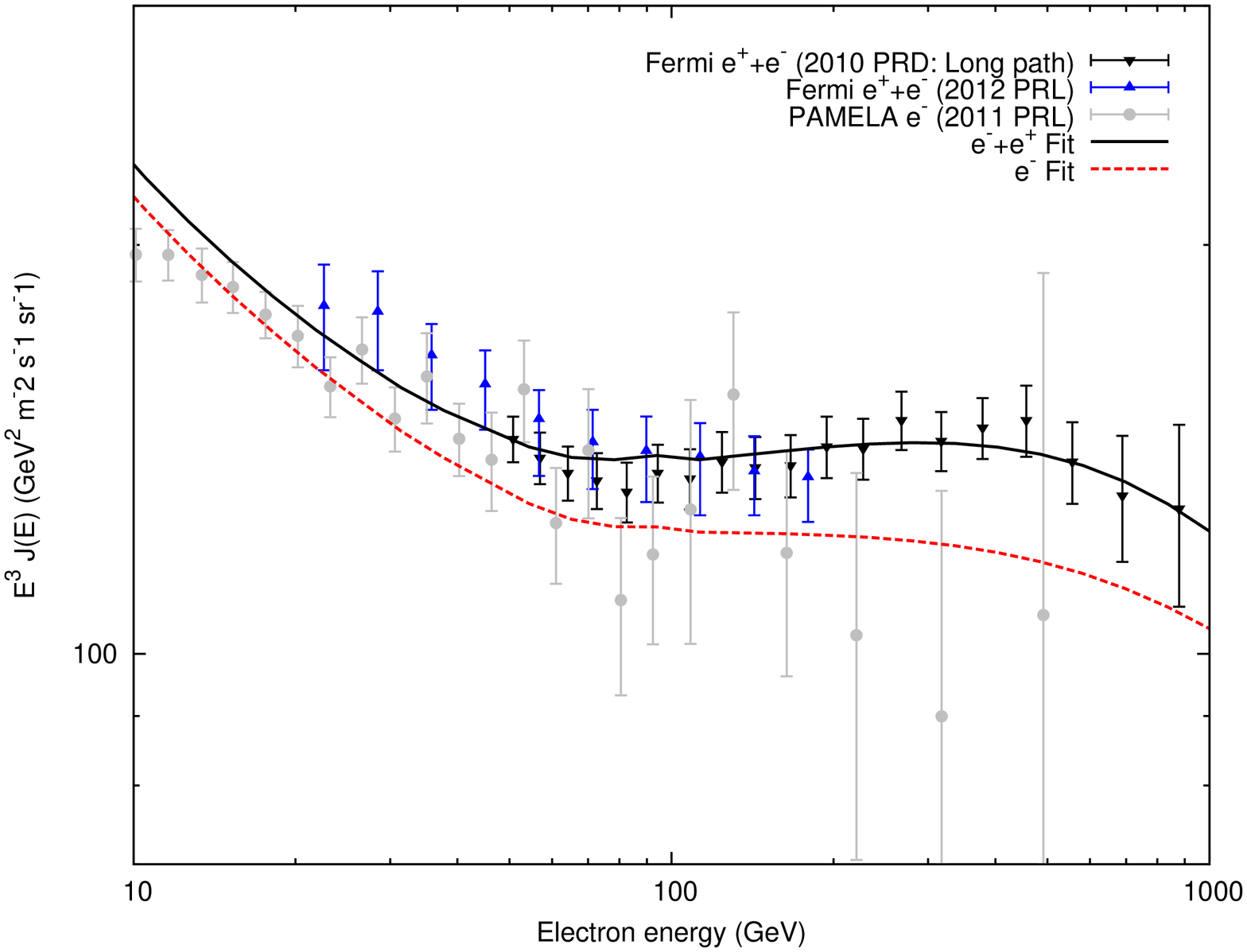}
\includegraphics[width=65mm,angle=270]{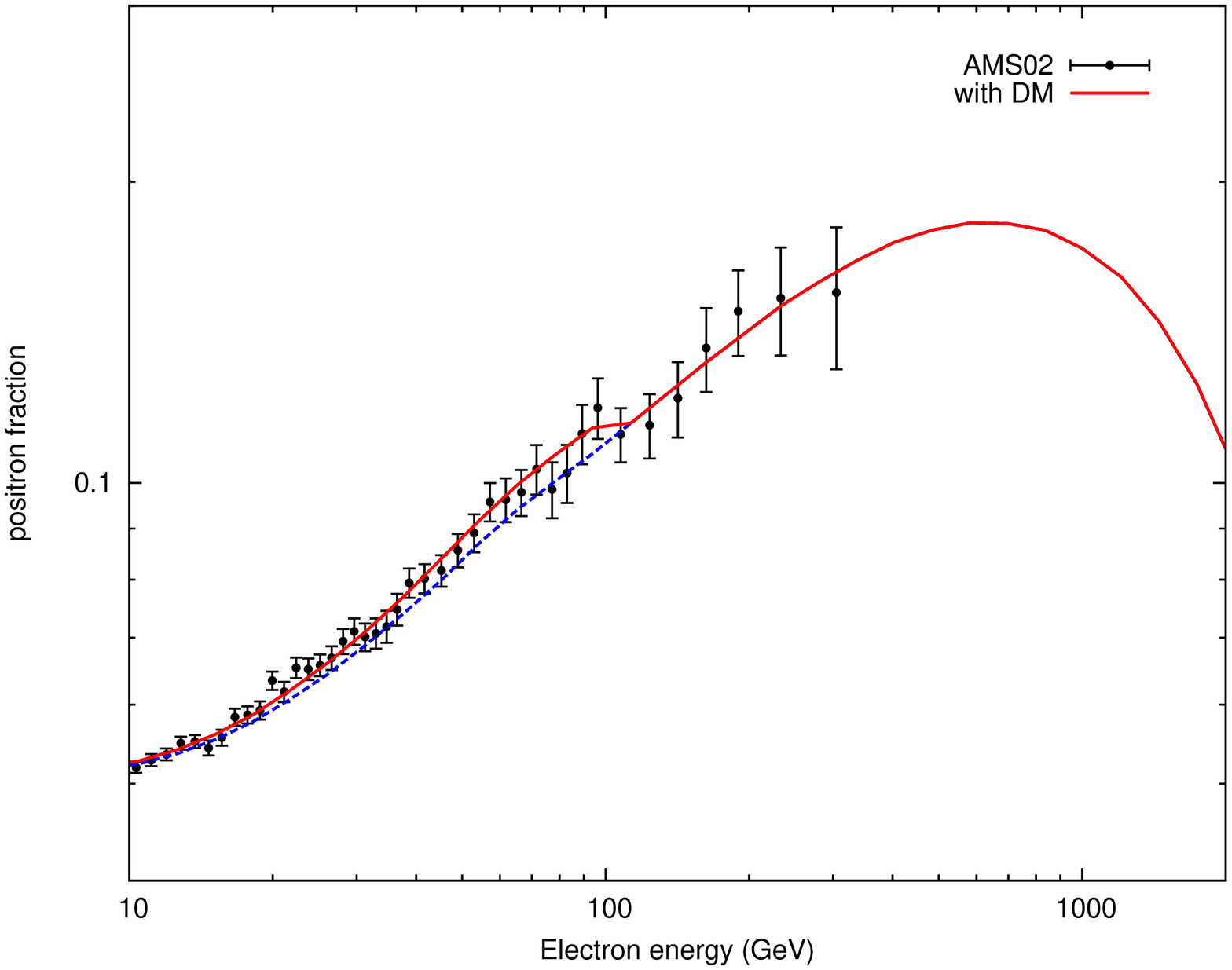}
  \caption{The ``background with spectrum hardening + pulsars + dark matter annihilation into $e^{+}e^{-}$" model for the second set of electron/positron data. Left panel: the fit of the Fermi-LAT long path electron/positron total spectrum \cite{fermi}, the updated Fermi-LAT electron/positron total spectrum in the energy range $20-200$ GeV \cite{Fermi2012}, and the PAMELA electron spectrum data \cite{Pamela2011}. Right panel: the fit of the AMS-02 positron fraction data. The best fit parameters are presented in Tab.\ref{table2} (i.e., scenario (e)). The existence of dark matter particles with a rest mass $\sim 100$ GeV and $<\sigma v>_{\chi\chi\rightarrow e^{+}e^{-}} \sim 5.5\times 10^{-27}~{\rm cm^3~s^{-1}}$ can not be ruled out by current data.}
  \label{fig:3}
\end{figure}

In Fig.\ref{fig:3} we discuss the possibility that the positron excess detected by PAMELA/Fermi-LAT/AMS-02 may be dominated by pulsar-like astrophysical sources and the contribution by dark matter may be small but detectable \cite{Feng2013}, i.e., scenario (e). The best fit of the second set of data yields an annihilation cross section $<\sigma v>_{\chi\chi\rightarrow e^{+}e^{-}} \sim 5\times 10^{-27}~{\rm cm^{3}~s^{-1}}$ and a dark matter rest mass $m_{\chi} \sim 104$ GeV. Currently the existence of such a component in the data can not be neither confirmed nor ruled out. More accurate data can help us to pin down such an issue. In particular, if the dark matter component is real, at the energy $\sim 104$ GeV the electron (positron) flux will drop by a factor of $\sim 1\%$ ($10\%$) suddenly, which can be directly tested by AMS-02 in the future.

\section{Conclusion and Discussion}
The Balloon-borne experiments ATIC and CREAM as well as the space-based experiment PAMELA
all show considerable cosmic-ray-nuclei spectrum hardening above the rigidity $\sim$ 240 GV. A natural question one would ask is whether
there is a corresponding spectrum hardening for primary cosmic ray electrons that can account for part of the $e^{-}+e^{+}$ excesses detected by Fermi-LAT, ATIC, HESS and PAMELA. The rigidity at which the possible electron spectrum hardening presents is hard to reliably estimate, so is the change of the spectral index of primary electrons. Speculatively, one can take for example $\epsilon_{\rm h,e}\sim 80~{\rm GV}$ and $\delta \sim 0.18-0.3$. With a given electron/positron total spectrum, the spectrum hardening of the primary electrons inevitably leads to the softening of the rest ``additional component" which has been widely assumed to consist of electron and positron pairs. Hence the increasing behavior of positron-to-electron ratio should be shallower than that in the absence of a spectrum hardening. A reliable estimate of the physical parameters of the ``additional component" emitter (from either dark matter annihilation/decay or pulsars) to address the electron/positron excesses is not achievable if the background spectrum hardening has not been properly taken into account. Our joint fits of the latest AMS-02 positron fraction data together with the PAMELA/Fermi-LAT electron/positron data suggest that the primary electron spectrum hardening is needed in most though not all modelings (see Tab.\ref{table1} and Tab.\ref{table2}). Such results are encouraging. However one should bear in mind that the data used in the modeling are from different instruments and may suffer from significant systematic errors. We are also aware that other models such as multiple pulsars and the decaying asymmetric dark matter can nicely reproduce the data. Therefore much more accurate data are still needed to test the spectrum hardening hypothesis further.

The bounds of the latest AMS-02 positron fraction data and the PAMELA/Fermi-LAT electron/positron spectrum data on dark matter models have also been investigated. In the presence of the spectrum hardening of primary electrons, the dark-matter-originated electron/positron pair component needed in the modeling is smaller since at $\epsilon>\epsilon_{\rm e,h}$ the flux of the hardened primary electrons is considerably larger. Even with such a modification, most of the parameter space for $\chi\chi \rightarrow \mu^{+}\mu^{-}$ has been excluded by the Fermi-LAT Galactic diffuse emission data. The decay channel $\chi\rightarrow \mu^{+}\mu^{-}$ is found to be viable. Alternatively, the positron excess detected by PAMELA/Fermi-LAT/AMS-02 may be dominated by pulsar-like astrophysical sources and the contribution by dark matter may be small but detectable. At energies $\leq 100$ GeV, a positron/electron component from either dark matter annihilation/decay or pulsar can not be ruled out (see Fig.\ref{fig:3} for illustration).

\acknowledgments This work was supported in part by 973 Program of China under grant 2013CB837000, by National Natural Science of China under grant 10925315, and by China Postdoctoral science foundation under grants 2012M521136 and 2012M521137.  YZF is also supported by the 100
Talents program of Chinese Academy of Sciences and the Foundation for
Distinguished Young Scholars of Jiangsu Province, China (No. BK2012047).
\\

$^\ast$Corresponding author.\\

\end{document}